# $PbTi_{1-x}Pd_xO_3$: A New Room-temperature Magnetoelectric Multiferroic Device Material

Elzbieta Gradauskaite,[1,§] Jonathan Gardner,[2] Rebecca M. Smith,[2] Finlay D. Morrison,[2] Stephen L. Lee,[1] Ram S. Katiyar,[3] and James F. Scott[1,2] [*]

[1] School of Physics and Astronomy, St. Andrews University, St. Andrews KY16 9SS, UK
[2] School of Chemistry, St. Andrews University, St. Andrews KY16 9ST, UK
[3] Dept. Physics, SPECLAB, Univ. Puerto Rico, San Juan, PR 90021 USA
[§] Present Address: Paul Scherrer Institut, 5232 Villigen, PSI, Switzerland.

**Abstract**

There have been a large number of papers on bismuth ferrite ($BiFeO_3$) over the past few years, trying to exploit its room-temperature magnetoelectric multiferroic properties. Although these are attractive, $BiFeO_3$ is not the ideal multiferroic, due to weak magnetization and the difficulty in limiting leakage currents. Thus there is an ongoing search for alternatives, including such materials as gallium ferrite ($GaFeO_3$). In the present work we report a comprehensive study of the perovskite $PbTi_{1-x}Pd_xO_3$ with $0 < x < 0.3$. Our study includes dielectric, impedance and magnetization measurements, conductivity analysis and study of crystallographic phases present in the samples with special attention paid to minor phases, identified as PdO, $PbPdO_2$, and $Pd_3Pb$. The work is remarkable in two ways: Pd is difficult to substitute into $ABO_3$ perovskite oxides (where it might be useful for catalysis), and Pd is magnetic under only unusual conditions (under strain or internal electric fields). The new material, as a PZT derivative, is expected to have much stronger piezoelectric properties than $BiFeO_3$.

*Author to whom correspondence to be addressed. Electronic mail: jfs4@st-andrews.ac.uk (J. F. Scott)

**Introduction**

   **1.1. Ferroelectric Memories**

Beginning around 1984 random access memories utilizing ferroelectric thin films (FRAMs or FeRAMs) were developed. In comparison with other non-volatile computer memories these were fast (60 ns in early embodiments), low-power (voltage-driven as opposed to current-driven magnetic memories), radiation-hard (no single-event upset SEU), and cheap.[1,2]

Within the past few years they have become a major disruptive industry at > $100 million/year, replacing magnetic-stripe cards and used for train and subway fares, cash-points ("e-money"), vending machines, employee identification cards, storage lockers, and convenience stores. They are contact-free cards, which can be updated (money added), and they are manufactured up to 8 Mb capacity by both Fujitsu and Toshiba in Japan and Texas Instruments in the USA (the latter for Washington subways). They are sold under the brand names FeliCa and Suica in Asia; Samsung has published 64 Mb FRAMs (the state of the art) but as far as we know is not marketing them commercially.[3]

   **1.2. Multiferroic magnetoelectrics**

In research laboratories worldwide one major effort has been to combine the advantages of FRAMs (speed, low power) with the advantages of magnetic devices (non-destructive read-out). This involves the use of materials that are simultaneously ferromagnetic and ferroelectric (multiferroic) and which have a linear (magnetoelectric) coupling between those two phenomena. $BiFeO_3$ has been the leading contender,[4,5,6,7] but other materials such as

GaFeO$_3$ have also been explored.[8] In the present paper we consider a new material, PbTi$_{1-x}$Pd$_x$O$_3$, that appears potentially superior to BiFeO$_3$ for some devices.

### 1.3. Pd in Perovskite Oxides

Pure perovskite-phase lead palladate PbPdO$_3$ is not stable (PbPdO$_2$ with Pd$^{2+}$ is favored), and indeed other investigators trying to substitute Pd into ABO$_3$ perovskite oxides report that the substitution is reversible and complicated due to various Pd oxidation states adopted.[9,10] However, we have found that up to 30% can be substituted into PbTiO$_3$, mostly at the Ti$^{4+}$ B-site, where it is a good match for valence and ionic size. We find that typically 30-40% of the Pd goes in as Pd$^{2+}$, however, a point to which we shall return.

In addition, Pd$^{4+}$ ([Kr] 4d$^6$, low-spin configuration) is normally not ferromagnetic; [11] however, it has an instability towards magnetism and can be ferromagnetic under internal electric fields or strain, [12,13] such as might be encountered if it is located not at the Ti$^{4+}$ B-site but at the oversized Pb$^{2+}$ perovskite A-site.

Thus in examining lead palladium titanate as a multiferroic, we are rejecting a popular view that it will not substitute in perovskite oxides and that it would not be ferromagnetic if it did. In this sense the present work leads to a radical new direction in multiferroics. We emphasize that although Pd is expensive (comparable to gold), the intended devices are thin films, typically 30-100 nm thick, for which the use of Pd is a negligible additional cost, and remind readers that Ag/Pd was the most common electrode material in ordinary multilayer capacitors (Kyocrera, TDX, AVX, etc.) until its relatively recent replacement by Ni.

## 1.4. Initial Experiments

It had been proposed that doping ferroelectric lead titanate ($PbTiO_3$) with palladium (Pd) results in a magnetoelectric multiferroic material at room temperature. This is remarkable because Pd atom itself is non-magnetic, but when it replaces lead (Pb) and titanium (Ti) ions, magnetism is observed as predicted from the density functional theory (DFT) studies[14] (discussed in section 1.5).

There is some experimental evidence for multiferroicity and giant magnetoelectric coupling at room temperature too. Kumari *et al*. [14] showed that 30% Pd doped lead titanate zirconate $Pb(Zr,Ti)O_3$ (PZT) displays room-temperature weak ferromagnetism, strong ferroelectricity, and strong ME coupling. However their work was not as conclusive as the present study, due to multiple unresolved secondary phases present in the samples and the rather complicated electronic structure of PZT.

The seminal work of Kumari *et al.* [14] is important but differs from the present in several important ways: (1) The main impurity phase $PbPdO_2$ could not be seen in XRD because its strong Bragg reflections are coincidentally under the main PZT peaks; (2) The magnetic properties measured did not reveal the magnetic $PbPdO_2$ contribution; (3) The SEM studies did not isolate any minor phases, including $PbPdO_2$; (4) Activation energies for conduction, important for device applications were unmeasured; and (5) No change in $T_c$ with Pd% could be measured due to the dependence of Curie temperatures on Ti/Zr ratio.

## 1.5. Predictive Theory

Ferromagnetism arising from doping PbTiO$_3$ with Pd was predicted by DFT calculations of Paudel and Tsymbal (Kumari *et al.* [14]) and more generally for Pd in earlier work by Sun, Burton, and Tsymbal [12].

Three different types of Pd doping were modeled computationally. When Pd$^{4+}$ substitutes Ti$^{4+}$ ion, the induced defect states lie in the conduction and valence bands with no occupied defect states in the band gap of PbTiO$_3$: isoelectric substitution does not produce magnetism. The same applies for Pd$^{2+}$ substituting for Pb$^{2+}$ ions. However, a magnetic moment of 2μ$_B$ is observed when Pd$^{4+}$ replaces both Ti$^{4+}$ and Pb$^{2+}$ cations that are first or second nearest neighbors. Exchange-split peaks are observed in the band gap of the system at 2.0 eV, related to spin polarization of both Ti and Pb density of states (DOS). These defect states hold two electrons that can be thought of as donated by a Pd$^{4+}$ ion residing nominally on the Pb$^{2+}$ atom. Most of the magnetic moment (1μ$_B$) is residing on the Pb site; Pd on the Ti site produces a small moment (0.1μ$_B$), and the rest comes from spin polarization of the DOS of oxygen atoms bonded to Pd replacing Pb.

**Experimental Details**

PbTiO$_3$ preparation: stoichiometric amounts of PbO and TiO$_2$ were calcined in PbTiO$_3$ sacrificial powder at 600°C and 800°C for 2h and 4h, respectively. Optimal sintering conditions were found to be 900°C for 16h. Pd-doped PbTiO$_3$ preparation: stoichiometric amounts of PbO, TiO$_2$ and PdO were calcined in sacrificial PbTiO$_3$ powder. B-site substitution only was assumed for stoichiometric calculations. Calcination temperatures were chosen between 600°C and 800°C for 1 to 10 hours. Pellets for electrical measurements were sintered at 700°C to 900°C for 4 to 16 hours to minimize the formation of Pd-containing side products. *PbPdO$_2$ preparation:* PbPdO$_2$ was prepared from stoichiometric amounts of PbO and PdO. Pellets were sintered in PbO sacrificial powder for 72 hours at 700°C.

PXRD measurements: PANalytical EMPYREAN powder X-ray diffractometer in reflection mode was used to collect XRD data. XRD patterns were collected for 2θ values of 10° to 90° with the measurement duration set to 1-2 hours. Rietveld refinement: All powder diffraction data were analyzed by the Rietveld method using the General Structure Analysis System (GSAS).[15] The cif model (crystallographic information file) of room temperature PbTiO$_3$ (*P4mm* symmetry)[16] modified by incorporation of Pd atoms as well as the cif models of PbPdO$_2$,[17] PdO[18] and Pd$_3$Pb[19] were used for refinements. Background and peak profiles were refined in order to determine parameters of the unit cell and weight fractions of crystallographic phases.

SEM and EDX: Micrographs were obtained using a Joel JSM 5600 scanning electron microscope set to a 5 kV accelerating voltage. All micrographs presented were obtained from the fracture surfaces of pellets. Oxford Inca EDX system was used for compositional analysis.

*XPS:* Pellets were mounted to the ultra-high vacuum (UHV) system using conductive tape. During spectra acquisition, flood gun was used at 2.0 eV. To remove surface layers of the pellets, Ar$^+$ ion bombardment was performed in the UHV system for 20 minutes at 44μA/3kV.

Dielectric and impedance studies: The pellets were electroded by applying conductive silver paint on both faces of the circular pellets. In some cases gold electrodes were applied using an Emitech K550x sputter-coater. A coating current of 30 mA was applied until the resistance measured across the electrode was under 50 Ohms. Dielectric measurements were performed using a Wayne Kerr 6500B impedance analyser: capacitance and dielectric loss (*tanδ*) values as were collected in the temperature range of 25°C to 550°C, with an applied AC excitation of 500 mV (frequencies ranging from 25 Hz to 2 MHz). Electroded pellets placed in a compression jig within a Carbolite MTF 10/25/130 benchtop tube furnace. The cooling/heating rates were 2 K min$^{-1}$. For the low temperature measurements (50 K - 300 K), an Agilent 429A impedance analyzer was used with the sample mounted in a closed cycle cryocooler.

Magnetization measurements*:* The SQUID measurements were made in Quantum Design MPMS SQUID magnetometer.

**Results and Discussion**

### 1.6. Scanning Electron Microscopy (SEM) and Energy Dispersive X-ray Spectroscopy (EDX) Measurements

SEM and EDX were combined to determine the ceramic microstructure and perform elemental analysis of the samples, particularly focusing on Pd content in different phases. SEM micrographs of various undoped and Pd doped $PbTiO_3$ samples reveal relatively small grains and a high-degree of porosity, which was expected as $PbTiO_3$ synthesis is known to be challenging. This is due to the volatilization of PbO at elevated temperatures, which often causes deviation from stoichiometry, inhomogeneous microstructure and porosity [20,21] and a high tetragonality (c/a = 1.064 with c = 0.4156 nm and a = 0.3902 nm[22]) that develops during cooling across the Curie point, which induces large internal stresses in the unit cell that can easily result in flaws and cracks. Sintering at low temperatures (600°C-800°C) results in the grain size of 0.7±0.2 µm, which increases to 2.7±0.6 µm when sintered at high temperature (900°C), which is illustrated in **Figure 1a**.

Pd does not easily substitute for cations in perovskite oxides. Even though it is used with perovskite oxides for catalysis, the perovskite acts as a supporting framework that allows dispersion and activation of metallic Pd nanoparticles at its surface. Substitution of B-sites is very rarely observed and normally is under 10%.[9,23,24] Therefore it was expected that during $PbTi_{1-x}Pd_xO_3$ preparation some other Pd-containing phases might. Powder X-ray diffraction (PXRD) patterns reveal impurity phases of $PbPdO_2$ and PdO in calcined powders of 10% and 30% Pd-doped lead titanate (see Figure S1, Figure S2 and Figure S3, Supplemental Material [25]). $Pd_3Pb$ phase was detected after sintering at 800°C or higher, therefore samples used for further measurements were prepared below this temperature.

These secondary phases were difficult to identify in the samples sintered at low temperatures, possibly due to their low concentrations and very small grain size (as shown in **Figure 1a**, left). They were successfully resolved only for samples with considerable amounts of impurity phases ($PbPdO_2$ and $Pd_3Pb$), prepared at high temperatures. **Figure 1b** shows the micrograph of different phases present in 30% Pd doped sample, which was sintered at 900°C. EDX data were used to assign phases to their chemical composition by considering elemental fractions of Pb, Ti and Pd. There appears to be a clear separation between Pd-containing $PbTiO_3$ phase (small circular grains, 1-2 µm) and undoped $PbTiO_3$ (large cuboids, 10 µm). The $PbPdO_2$ phase was identified too, as a spherical cluster of very small grains.

Samples analyzed can be grouped into undoped and doped samples, which were further grouped depending on their sintering properties. The variations in particle size and effective Pd substitution for each group are shown in **Table 1**. Increase in particle size was observed not only at higher sintering temperatures but also after sintering in oxygen atmosphere. EDX studies helped to determine that Pd substitutes better at low temperatures. Synthesis under oxygen flow at low temperatures is optimal as it maximizes both ferroelectric (related to grain size) and ferromagnetic properties (related to higher Pd fraction), the only drawback being inhomogeneity of the ceramics due to an oxygen vacancy gradient.

### 1.7. X-ray Photoelectron Spectroscopy (XPS)

XPS data were collected from the surface of 30% Pd-doped pellet. Pd $3d_{5/2}$ and Pd $3d_{3/2}$ regions were detected at binding energies of 336.95 eV and 342.07 eV, respectively (**Figure 2a**). The peaks are symmetric and coincide with the region expected for $Pd^{2+}$. Therefore it was concluded that only $Pd^{2+}$ cations are present on the surface of the pellet. The observation can be explained by oxygen loss from the surface, which turns $Pd^{4+}$ into $Pd^{2+}$.

To investigate bulk of the pellet, $Ar^+$ bombardment in the UHV setup was performed to remove the surface of 10% Pd doped $PbTiO_3$. The Pd $3d_{5/2}$ and Pd $3d_{3/2}$ regions were no longer symmetric, revealing presence of both $Pd^{4+}$ and $Pd^{2+}$ states, required for magnetism (**Figure 2b**). The Pd $3d_{5/2}$ and Pd $3d_{3/2}$ doublets were deconvoluted into $Pd^{2+}$ components at 336.82 eV, 342.15 eV, and $Pd^{4+}$ at 337.72 eV, 343.05 eV.

A shift observed in $Pd^{2+}$ binding energy can be attributed to differential charging effects induced by ion beam. The ratio of $Pd^{4+}$ to $Pd^{2+}$ states was determined to be 1.47:1.00. According to the DFT calculations, $Pd^{4+}$ is required to be present in $PbTiO_3$ in order for magnetism to be observed - this condition is fulfilled.

### 1.8. Dielectric Studies

The dielectric spectroscopy data collected for undoped $PbTiO_3$, as well as 10% and 30% Pd doped pellets are shown in **Figure 3a**. The profile of the dielectric curves reveal one peak corresponding to ferroelectric to paraelectric transition. All three samples were also found to exhibit frequency independent (normal ferroelectric) behavior (**Figure 3b**), which is useful for device applications.

The magnitude of relative permittivity was strongly influenced by the quality of ceramics analyzed: Pd doping improved densification of pellets (density increased up to 40%); dense ceramics with fewer pores were observed to have larger relative permittivity values throughout the temperature sweep. This is normally explained by 90° domain size effects, which contribute for a very large part of relative permittivity.[26] The number of ferroelectric 90° domains per volume is maximized for well-sintered ceramics. Another explanation could possibly be the strain induced by Pd doping. Due to the mismatch in radii of Ti and Pd (0.01Å [27]), the volume of unit cell slightly increases upon doping. Therefore undoped nanoregions expand under the stress of the matrix,[28] which creates more space between Ti atoms and the oxygen octahedra allowing larger off-center displacements directly related to polarization strength. [28,29]

Upon Pd doping, the measured ferroelectric $T_c$ values were lower. This can be rationalized by changes in the unit cell induced by doping with $Pd^{4+}$ ion, whose radius (0.615 Å [27]) is slightly larger than that of $Ti^{4+}$ (0.605Å [27]). The phenomenon of transition temperature decreasing with the substitution of larger cations for smaller cations was observed in various perovskites and related structures and the effect can be as large as a 100°C shift [30,31].

The relative permittivity profile in 30% Pd doped samples is rather different from that observed for undoped and 10% Pd doped $PbTiO_3$. Even though in the measured temperature range there is only one transition, above $T_c$ the relative permittivity continues to increase. This is anomaly observed due to conduction processes in the ceramic, when conducting electrons and/or ions are measured as displacement current related to surface polarization charge density. Conduction is commonly observed in lead-containing ferroelectrics and is attributed to oxygen vacancies [32,33].

Oxygen vacancies become mobile at elevated temperatures (> 150°C [34]) and ionic conductivity becomes appreciable. However, oxygen vacancies cannot be the only source of conductivity in Pd-doped $PbTiO_3$. The 10% Pd doped sample sintered at 900°C for 16h (in which lead loss should be maximal) does not show anomalous increase in dielectric constant, whereas 30% Pd doped pellets sintered at much lower temperatures and for shorter durations exhibit considerable conductivity. This observation reveals that conductivity is mainly created by $PbPdO_2$ impurity phase (larger weight-fraction in 30% doped sample, see Figure S3, Supplemental Material [25]). This is not surprising as $PbPdO_2$ is reported to have a metallic-like conductivity at 90 K - 300 K. [17]

Kumari *et al.*[14] have shown that PZT ($PbZr_{1-x}Ti_xO_3$) is ferroelectric and magnetoelectric multiferroic at room temperature over a wide range of Zr/Ti ratios, using the positive-up negative-down (PUND) method and quantitative measurements of magnetoelectric tensor components, so we do not show such data here. We do find that leakage current is a problem for 30% Pd at T = 292K and suggest that such studies will be much more precise with thin films rather than our bulk ceramic specimens. Such films have not yet been made but should be possible with spin-on techniques and a liquid Pd source, or via sputtering.

### 1.9. Impedance Spectroscopy

Impedance spectroscopy measurements were performed on undoped, 10% and 30% Pd doped $PbTiO_3$. All plots show only one feature: semicircles in the complex plane modulus and impedance plots are symmetric, while Debye peaks in Z" and M" spectroscopic plots have nearly the same peak frequency, indicating they originate from the same single electroactive region (impedance plots collected for 10% Pd doped $PbTiO_3$ are shown in **Figure 4)**.

A single electroactive region observed in impedance spectra normally corresponds to a very well sintered sample with very large grains and thin and well-defined grain boundaries. On the contrary, SEM studies of both undoped and Pd doped $PbTiO_3$ samples show very small grains. Therefore it is very unlikely that grain boundary contribution is negligible – electroactive regions probably have similar time constants $\tau = RC$.

Nyquist semicircles in M* plot are distorted in the high frequency range. This can be explained by inductance of the measuring leads at high frequencies. Furthermore, Pd-doped $PbTiO_3$ samples are magnetic which also amplifies this effect. The parasitic inductance also affects semicircles in Z* plots: it causes a tail at high frequencies, which crosses the real impedance axis, but this is not obvious in the data collected.[35]

### 1.10. Conductivity Analysis

The control of conductivity is essential in ferroelectric ceramics - they should be insulating to minimize dielectric loss. The bulk conductivity in the three pellets and their variation with temperature was extracted from complex impedance and electric modulus measurements ($M''$, $M^*$, $Z''$ and $Z^*$ values). Ferroelectric materials are normally wide-band gap semiconductors and their temperature dependent conductivity processes generally follow the Arrhenius law: $\sigma = \sigma_0 \exp\left(-\frac{E_a}{kT}\right)$, where $\sigma$ is the conductivity, $\sigma_0$ is a pre-exponential factor and $E_a$ is the activation energy.

Arrhenius plots and corresponding parameters (**Figure 4**) reveal different features of electrical transport in undoped, 10% and 30% Pd doped PbTiO$_3$ pellets. PbTiO$_3$ exhibits two resolved thermally activated conductivity regions - in ferroelectric and in paraelectric regions. The activation energy is similar for both (0.58 eV and 0.46 eV, respectively), which suggests the same type of conduction mechanism. For the 30% Pd samples, the 0.74±0.02 eV activation energy most likely arises oxygen vacancy hopping; it agrees with the literature value of 0.74 eV in the paraelectric phase of PZT.[34] Migration barriers for oxygen vacancies are known to be phase dependent,[36] which explains the slight difference in activation energy values. The 10% doped pellet exhibits similarly behaved conductivity in paraelectric region, however, the conductivity in ferroelectric region is no longer resolved. This can be explained by PbPdO$_2$ phase, conductivity of which decreases with temperature and is dominant at low temperatures,[17] hindering Arrhenius behavior of oxygen vacancies, which becomes dominant at high temperatures. The conductivity in the 30% Pd doped sample is more complex. The sample is highly conducting even at room temperature and therefore little change is observed at low temperatures. A thermally activated conductivity region is resolved below T$_c$ with E$_a$=0.73 eV, which is not too different from oxygen vacancy conduction observed previously. Conductivity in the paraelectric phase has a very low $E_a$ of only 0.335 eV, which is characteristic of electronic conduction or mixed ionic-electronic conduction.

### 1.11. Effect of Oxygen Annealing

To improve dielectric properties of Pd-doped $PbTiO_3$, pellets were sintered in a flowing oxygen atmosphere. This oxygen annealing was expected to decrease oxygen vacancy concentration and the amount of $PbPdO_2$ present in the samples. The relative permittivity profile of the 30% Pd-doped $PbTiO_3$ is significantly improved by oxygen annealing (**Figure 5a**). Comparison of relative permittivities shows that oxygen annealing suppresses conductivity and also promotes Pd substitution into perovskite structure as previously supported by EDX studies (decrease in $T_c$). Comparison of the XRD patterns (**Figure 5b**) shows that sintering in oxygen atmosphere decreases the amount of $PbPdO_2$, with a minor increase in PdO concentration.

### 1.12. Magnetization Studies

Magnetization measurements were performed as a function of magnetic field on samples of both 10% and 30% Pd doped $PbTiO_3$ samples, as well as on a separately prepared sample of the main magnetic impurity phase $PbPdO_2$ (**Figure 8a**). The data on all samples were found to exhibit a significant diamagnetic contribution to the signal, clearly evident over most of the temperature range. In the plots presented, this contribution has been subtracted by performing a linear fit to the raw data in order to better highlight the ferromagnetic contributions of interest.

The 10% and 30% Pd doped samples are found to be weakly ferromagnetic even at room temperature (**Figure 8b**), with 30% Pd doped sample possessing a significantly larger magnetic moment at saturation ($0.2 \cdot 10^{-4}$ and $2.8 \cdot 10^{-4}$ emu g$^{-1}$ respectively). Both samples also exhibited an appreciable hysteresis, with the highest coercivity being found for 10% doped sample (~260 Oe at 300K). For device applications a high coercivity is an essential requirement, as it determines the stability of the memory by making it more robust to the influence of external magnetic fields. Further measurements made at 400K on 10% Pd doped PbTiO$_3$ still indicated strong ferromagnetic behavior.

Since both 10% and 30% Pd doped PbTiO$_3$ samples contain measurable fractions of PbPdO$_2$, it is imperative to ascertain the possible contributions of this impurity phase to the magnetic signals. PbPdO$_2$ is known to be diamagnetic at temperatures higher than 90 K, [17, 37, 38] but the sample measured here was ferromagnetic at 300 K (inset Fig 10a). The character of this phase is however very different from that observed in the perovskite samples, with a very small coercivity (~ 50 Oe) and a much higher saturated moment (Figure S4, Supplemental Material [25]). It is known that when doped with magnetic metal ions, such as iron, [39] copper or cobalt [38], that PbPdO$_2$ can become ferromagnetic even at room temperature, and it is likely that this is the origin of the ferromagnetic signal in the PbPdO$_2$ measured here. While the significantly higher moment of PbPdO$_2$ means that only a small amount of this impurity phase, if similarly doped, might make measureable ferromagnetic contributions to the signals observed for 10% and 30% Pd doped samples, the very different character of the ferromagnetic signal in both cases confirms that the ferromagnetism observed in the Pd-doped PT samples is intrinsic to those materials.

To gain some insight into the evolution of the magnetic order with temperature, a saturation field was applied at 2K and then the field set to zero so that the remanent magnetization could be recorded. The sample was then heated in zero field and the value of the magnetization was recorded as a function of temperature (Figure 11a). In this way any changes in magnetic order are likely to manifest as rapid changes in magnetisation. The general trend for both 10% and 30% Pd doped $PbTiO_3$ is a slow reduction of magnetization with increasing temperature, as would be expected for an ordered system. There is however a small feature around 90K. This does not appear to correlate with any features found in relative permittivity measurements, and in fact occurs at the known metal-insulator transition (90K) for undoped $PbPdO_2$.[17] This feature thus seems most likely to arise from a change in conductivity of the impurity phase and not from any changes to the perovskite phase. It is thus reasonable to conclude that the magnetic state of the multiferroic samples remains essentially unchanged from 2K up to the highest temperature measured (400 K).

**Conclusion**

Pd doped $PbTiO_3$ perovskite was found to be with Pd substitutional at both Pb- and Ti-sites and was shown to exhibit ferromagnetic properties at temperatures as high as 400 K. High levels of Pd substitution (30%) were observed in the perovskite structure along with the formation of $PbPdO_2$ and $Pd_3Pb$ impurity phases. $Pd^{4+}$ and $Pd^{2+}$ states were identified, fulfilling the condition for ferromagnetism determined from the density functional theory studies by Kumari *et al.*[14] Lowering of ferroelectric Curie point (to 740 K and 730 K for 10% and 30% doping, respectively) and increase in relative permittivity (by 1.5 times and 7.0 times for 10% and 30% doping, respectively) were observed. The samples exhibited relatively low saturation magnetisation at room temperature ($0.2 \cdot 10^{-4}$ and $2.8 \cdot 10^{-4}$ emu g$^{-1}$ for 10% and 30% doping, respectively), but high coercive field (258 Oe). Even though a ferromagnetic signal was detected for a minor phase present in the sample (PbPdO$_2$), it was shown that it is a small contribution to the total signal. Pd-doped samples showed increase in conductivity, which was explained by metallic-like conductivity of PbPdO$_2$ in the lower temperature range and oxygen vacancy, as well as electronic conductivity, at high temperatures. Oxygen annealing was shown to be effective in decreasing this lossy behavior.

In order to produce prototype room-temperature devices it is still necessary to eliminate minor phases, especially PbPdO$_2$ and Pd$_3$Pb, which occur in our ceramic specimens and exacerbate the leakage currents. New efforts are suggested for thin-film fabrication, including sol-gel spin-on and sputtering techniques. However, the present work serves as an existence proof for compounds that will combine the excellent piezoelectric and pyroelectric properties of PZT with ferromagnetism. This has obvious commercial applications for nonvolatile memories as well as for transducers and actuators. In comparison with two-component multiferroic sandwich devices, which are limited by strain coupling and hence the speed of sound, they offer improved response time.

**Acknowledgements**
A part of the work was carried out at the University of Puerto Rico (UPR) with financial support provided by the DoD-AFOSR Grant # FA95501610295. J.F.S. acknowledges his visit expenses to UPR from IFN-NSF Grant # 1002410. Work at St. Andrews was supported by EPSRC grant EP/P024637/1.

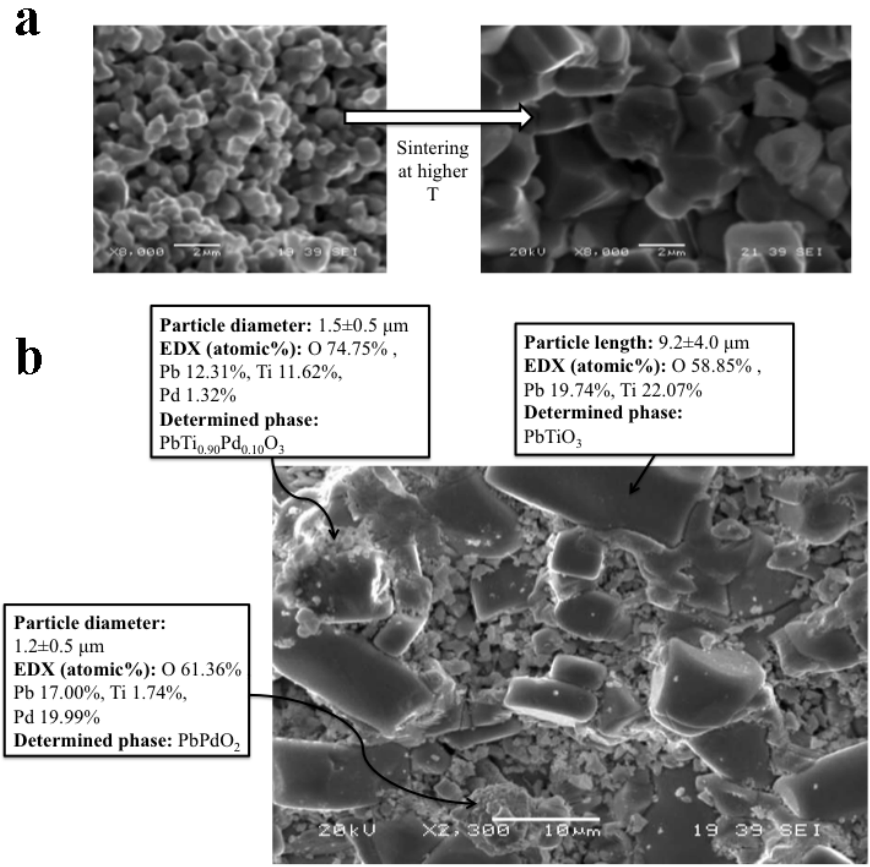

**Figure 1.** a) SEM micrographs showing the effect of sintering at higher temperature: micrographs show samples sintered at 750°C (left) and 900°C (right). b) SEM micrograph of 30% Pd doped sample, sintered at 900°C. Elemental composition of different phases and particle sizes are shown.

**Table 1.** Particle size and effective Pd substitution to perovskite phase: variation with different synthesis conditions.

| Group | Sintering T (°C) | Pd doped (%) | Particle Size (μm) | Detected Pd (%) |
|---|---|---|---|---|
| Undoped PbTiO$_3$ | 700 | 0 | 0.7 ± 0.2 | nil |
| | 900 | 0 | 0.6 ± 0.2 | |
| Sintered at low T | 750 | 10 | 0.8 ± 0.3 | 10 |
| | 750 | 30 | 0.7 ± 0.1 | 26 |
| Annealed in O$_2$ | 750 | 10 | 2.4 ± 1.7 | 25 and 9 |
| | 750 | 10 | 1.6 ± 0.3 | 10 and 3 |
| Sintered at high T | 900 | 10 | 2.7 ± 0.6 | 8 |
| | 900 | 30 | 1.5 ± 0.5 | 10 |

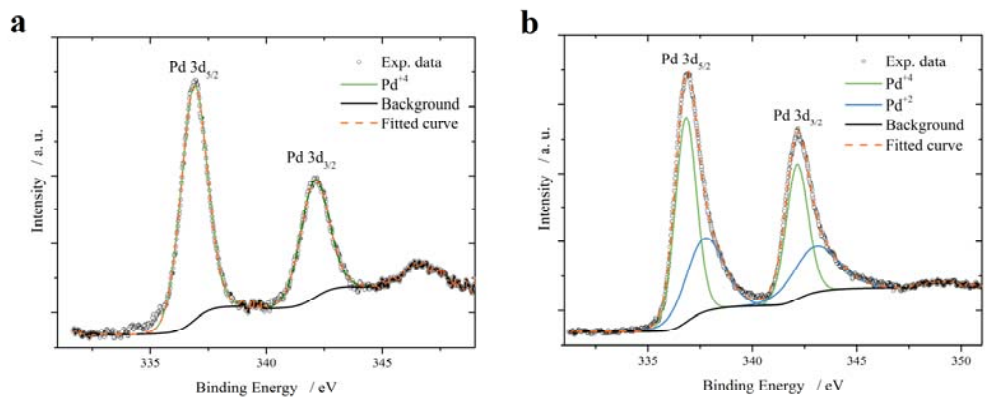

**Figure 2.** XPS spectra of the Pd 3d peaks from the surface of 10% Pd doped PbTiO$_3$ pellet: a) before Ar$^+$ bombardment; b) after Ar$^+$ bombardment.

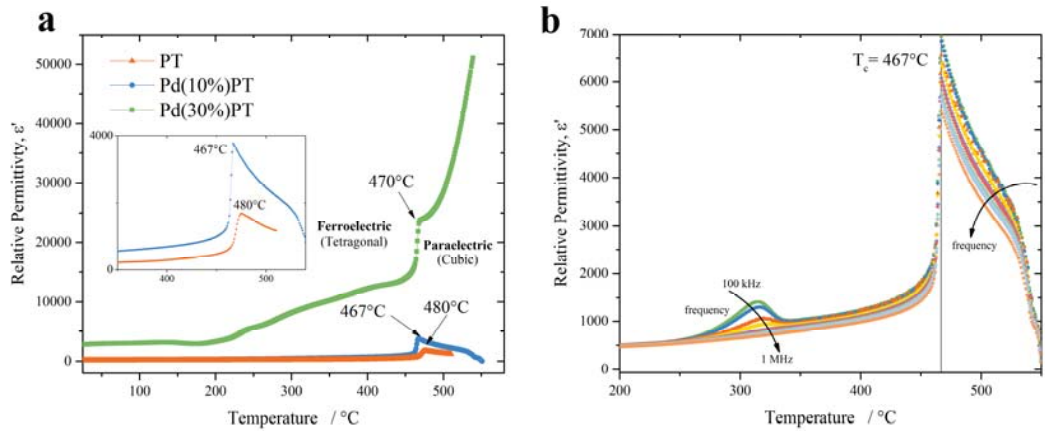

**Figure 3.** a) Relative permittivity versus temperature for undoped, 10% and 30% Pd doped PbTiO$_3$ pellets. T$_c$ values are labeled. b) Relative permittivity of 10% Pd doped PbTiO$_3$ as a function of temperature and frequency (feature at 325°C corresponds to the curing of silver electrodes).

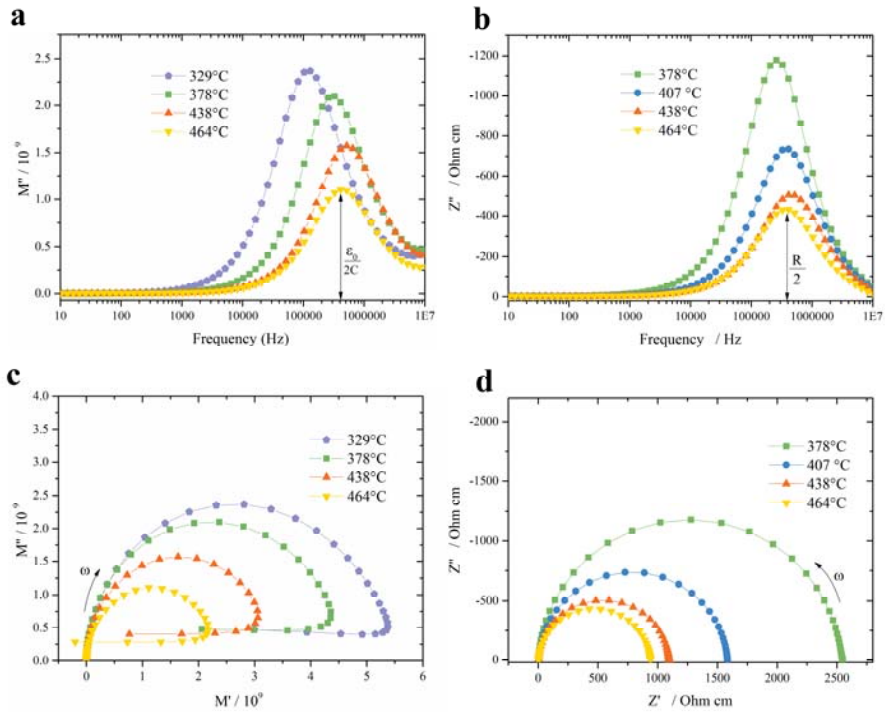

**Figure 4.** High-temperature impedance spectroscopy data collected for 10% Pd doped PbTiO$_3$ sample: a) M'' spectroscopic plot; b) Z'' spectroscopic plot; c) complex plane M* plot and d) complex plane Z* plot.

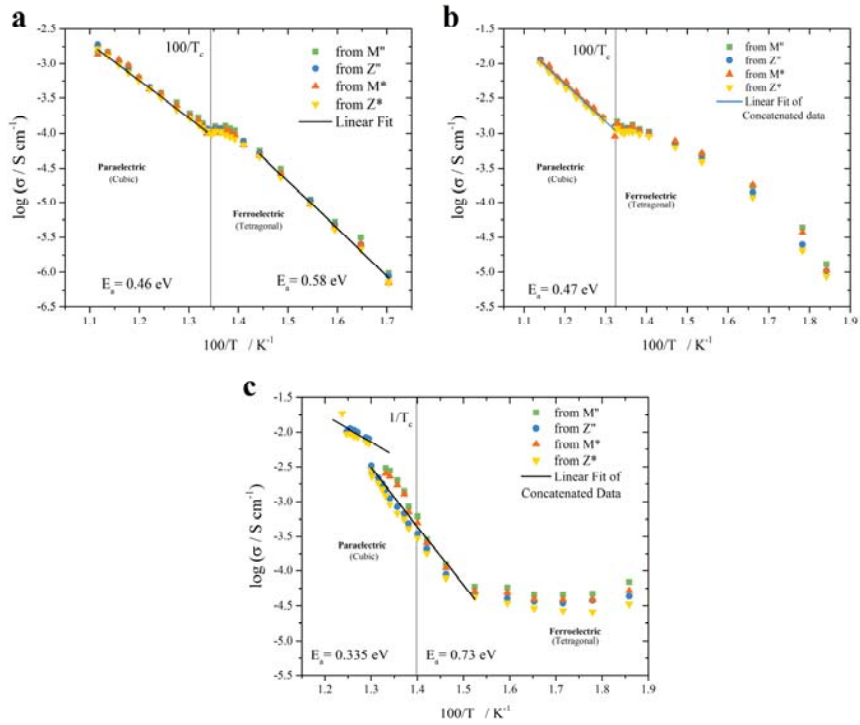

**Figure 5.** Plots showing temperature dependence of conductivity for (a) undoped, (b) 10% Pd-doped and (c) 30% Pd-doped $PbTiO_3$. Data points were extracted from values measured for M", M*, Z" and Z*.

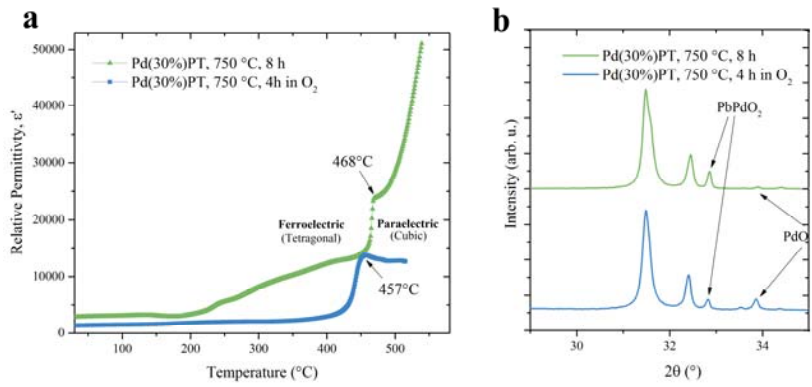

**Figure 6.** The oxygen annealing effect on 30% Pd doped PbTiO$_3$: (a) difference in relative permittivity and (b) XRD patterns when compared with sintering in ambient atmosphere.

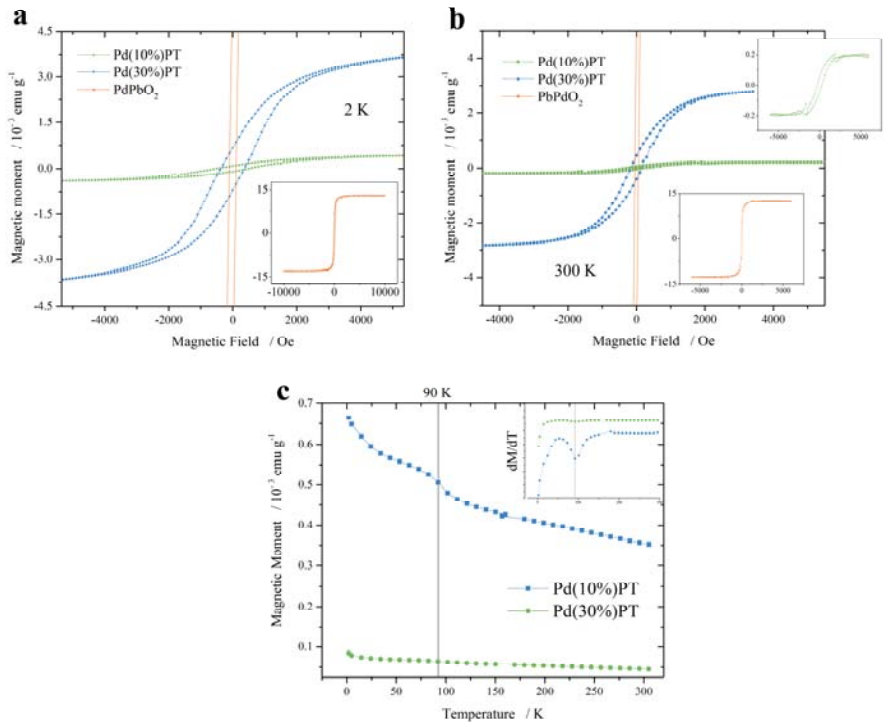

**Figure 7.** Comparison of magnetisation M(H) for 10%, 30% Pd doped $PbTiO_3$ and the $PbPdO_2$ secondary phase at 2K (a) and 300K (b). Insets show M(H) for $PbPdO_2$ (a,b) and 10% doped $PbTiO_3$ (b). (c) Evolution with temperature of the remanent magnetisation created at 2 K, for 10% and 30% Pd doped $PbTiO_3$. The inset shows variation in magnetization derivative with temperature.